\documentclass[12pt,english]{article}
\usepackage[T1]{fontenc}
\usepackage[latin1]{inputenc}
\usepackage{graphicx}
\usepackage{setspace}
\makeatletter

 \newcommand{\lyxaddress}[1]{
   \par {\raggedright #1 
   \vspace{1.4em}
   \noindent\par}
 }

\usepackage{babel}
\makeatother
\begin{document}

\title{Relating Calculations and Renormalization in Axial and Lorentz Gauges
and Gauge-independence }

\author{Satish D. Joglekar}

\maketitle

\lyxaddress{{\large Department of Physics, Indian Institute Of Technology, Kanpur,
Kanpur 208016 {[}INDIA{]}}}

\begin{abstract}
We study futher the recently developed formalism for the axial gauges
toward the comparison of calculations and of the renormalization procedure
in the axial and the Lorentz gauges. We do this in the 1-loop approximation
for the wavefunction renormalization and the identity of the $\beta$-functions
in the two gauges. We take as the starting point the relation between
the Green's functions in the two gauges obtained earlier. We obtain
the relation between the 1-loop propagators in the two gauges and
locate those diagrams that contribute to the difference between the
wave-function renormalizations in the two gauges. We further employ
this relation between the Green's functions to the case of the 3-point
function and prove the identity of the beta functions in the two gauges.
\end{abstract}

\section{Introduction}

Calculations in the standard model have been done in a variety of
gauges\cite{ijp}. Among these, the two of the most commonly used
are the Lorentz and the axial gauges \cite{bns,l}. The axial gauges
have generally been treated so far by means of a prescription of some
kind \cite{bns,l} and the contact with the results of the Lorentz
gauges is established only at the end of the calculation of a given
physical observable. Thus, so far the procedure used during the intermediate
stages of the calculations and of renormalizations, that have been
carried out in these two gauges, has been totally unrelated. The veracity
(or otherwise) of the gauge-independence of the results for observables
(and therefore the correctness of prescription) is known only at the
end. In fact, in literature, discrepancies have occasionally been
reported between the results of the two gauges. 

Recently, we have emphasized \cite{j,011} the need to carefully deal
with the boundary condition term (the $\epsilon-term$) while defining
the non-covariant gauges via interpolating gauges. We have also emphasized
\cite{mpla02,mpla03} the delicate and, \emph{a priori} risky nature
of the path-integral for the non-covariant gauges%
\footnote{It should be emphasized that the results in \cite{mpla02,mpla03}
are not necessarily to be thought of as being due to illegitimate
manipulations of these path-integrals. There is a correlation between
the Feynman rules approach to QFT and the Lagrangian path-integral
approach and these problems could be translated in the language of
the Feynman rules approach.%
}. We have also seen there that imposing a prescription \emph{by hand}
may not necessarily be compatible with the symmetries of the problem.
These point to a need for an additional care while formulating non-covariant
gauges.

The author, with a number of coworkers (Misra, Mandal, Bandhu), has
recently developed a formalism for the non-covariant gauges that enables
one to construct the non-covariant gauge path integral that is compatible%
\footnote{By this, we mean that the expectation value of \emph{any gauge-invariant
operator} is the same for both the gauges (made already well-defined)
in this formalism. If one imposes a prescription by hand, this equality
may not generally hold: recall the case of CPV in temporal gauge \cite{ccm}.%
} by construction with Lorentz gauges \cite{jmp00,ijmpa00} and does
not impose a prescription by hand. In this formalism, one takes care
of the boundary condition term correctly in an \emph{intrinsic} fashion.
Such a path integral directly takes care of the axial pole problems
and has been employed to derive the correct treatment of the axial
poles \cite{ijmpa00,mpla99}. It has also been employed to address
to the problems associated with the Coulomb gauge \cite{jm-02}. In
this formalism, it is possible to relate the Green's functions in
the two gauges by a simple relation \cite{jmp00,ijmpa00} (See Eq.(\ref{basic})).
This relation, obtained via a field transformation in the BRS space
\cite{jm95,bj98}, intrinsically incorporates the correct way of handling
the axial poles \cite{ijmpa00,mpla99} and preserves the expectation
value of all gauge invariant observables \cite{jmp00,jm95}. In this
work, we further explore applications of this result for the renormalization
procedure in these two gauges. 

We would like to compare this work with the earlier works on the renormalization
of gauge theories in the axial-like (including the light-cone) gauges.
For the light-cone gauge, results regarding the counter-term structure
(together with the problem of non-local divergences) have been worked
at for the Leibbrandt-Mandelstam prescription \cite{bds87}. For the
axial gauges other than the light-cone gauge $(\textrm{$\eta^{2}$}\neq0)$,
results regarding the (local) nature of counter-terms have been developed
for the CPV (Cauchy Principal Value prescription) to all orders of
perturbation theory \cite{bns}. However, as has been known, the axial
gauges defined with CPV do not necessarily yield results consistent
with the Lorentz gauges \cite{ccm,bns}%
\footnote{That such a possibility is always imminent when a prescription is
imposed should be apparent, in particular, from the conclusions of
\cite{mpla02,mpla03}.%
}. Attempts have been made to develop results related to the {}``uniform
prescription'' which is a prescription constructed for the axial
gauges by analogy with the Leibbrandt-Mandelstam (LM) prescription
for the light-cone gauges \cite{bns,l}. Like the LM prescription,
it has non-local divergences. It has been developed upto 1-loop approximation. 

In this work, we shall address to some of the various questions that
will help make this formalism useful in additional directions. Among
the possible applications of this formalism would be to develop techniques
that will (i) correlate the calculations in the two gauges (ii) correlate
the renormalization programs in the two gauges, and thus (iii) develop
theoretical techniques to establish the gauge-independence of observables.
It also remains to establish a renormalization procedure for the axial
gauges so constructed. One of the essential results necessary in this
direction is the investigation of the nature of divergences in these
gauges. We have made a preliminary investigation of this at one-loop
level and found only local divergences \cite{jm-01} in the propagator.
We shall find this result useful in this work. Unlike the uniform
gauges and the Leibbrandt-Mandelstam prescription for light-cone gauges
(which are {}``2-vector'' prescriptions \cite{bns,l}), we could
expect only local divergences. We shall leave the issue to a future
work.

Suppose we work in a given class of gauges, say the Lorentz gauges
with a variable gauge parameter $\lambda$ . We can, then, relate
the gauge-dependence of the Green's functions and of the renormalization
procedure (and in particular of the renormalization constants) for
gauge parameters differing infinitesimally. This procedure can be
established based on the early works on the renormalization of gauge
theories \cite{lh,Niel}. We show that an identical procedure can
be developed for comparing renormalization procedures in totally different
classes of gauges such as the Lorentz and the axial gauges. (However,
there are major technical differences/difficulties in the present
case arising from properties of the propagator denominators and regarding
power counting etc). As an example, we shall establish this connection
for the renormalization constant for the gauge field and for the beta
function both in the one loop approximation. We show, in particular,
that the renormalization constants in the two gauges for the gauge
field can be related by a similar procedure as was possible for two
sets of gauges differing infinitesimally within a given class of gauges.
In particular, we show {[}without explicit calculation{]} the result
that the one loop beta function for the axial and Lorentz gauges in
the minimal subtraction scheme are identical. Our results indicate
that it is generally possible to compare the renormalization procedures
and observables in two distinct classes of gauges with comparative
ease. 

We shall briefly state the plan of the paper. In section 2, we shall
review the formulation of references \cite{jmp00,ijmpa00} that relates
the Green's functions in the two gauges. In section 3, we shall apply
this relation to obtain the relation between the 1-loop propagators
in the two gauges and locate those diagrams that contribute to the
difference between the wave-function renormalizations in the two gauges.
In section 4, we shall apply the relation between the Green's functions
to the case of the 3-point function and prove the identity of the
beta functions in the two gauges.

\section{{\normalsize Preliminaries}}

\subsection{The Formulation}

We work in the axial-type gauges with a gauge parameter $\lambda$
and \[
S_{eff}^{A}=-\frac{1}{2\lambda}\int d^{4}x(\eta\cdot A)^{2}\]
and regard $\eta\cdot A=0$ as the $\lambda\longrightarrow0$ limit
of the above family of gauges. We also have the ghost action:\[
S_{gh}^{A}=-\int d^{4}x\,\bar{c}\eta^{\mu}D_{\mu}c\]
In the references \cite{jmp00,ijmpa00}, we established a procedure
for correlating the path integral for the axial gauges to that for
the Lorentz gauges using a field transformation called the FF-BRS
transformation \cite{jm95,bj98}: a generalization of the BRS transformation.
We noted that the proper definition of Green's functions in axial
gauges is not possible unless the procedure for the treatment of the
$\frac{1}{\eta.q}$-type singularities is first established. In the
references \cite{ijmpa00,jmp00}, we gave a procedure for doing this
in a manner compatible with the Lorentz gauges.

\paragraph{\textmd{The proper definition of the Green's functions in the (covariant)
Lorentz gauges requires that we include a term $-i\epsilon\int d^{4}x(\frac{1}{2}A^{2}-\overline{c}c)$
in the effective action that, in particular, determines the boundary
conditions to be associated with the unphysical degrees of freedom.
Similarly, proper definition of the axial Green's functions in Axial-type
gauges requires that we include an appropriate $\epsilon$-term.The
correct $\epsilon$-term in the axial-type gauges can be obtained
by the FF-BRS transformation that connects the path integrals in the
Lorentz and the axial gauges as was done in} \textmd{\cite{ijmpa00,jmp00}.}
\textmd{We gave two alternate forms for the relation between Green's
functions in the two sets of gauges \cite{jmp00}. It was shown \cite{ijmpa00}
that the effect of this term on the axial gauge Green's functions
is expressed in the simplest form when it is expressed in the relation
(\ref{basic}) given below and happens to be simply to add to $S_{eff}^{M}(A,c,\overline{c})$
the same $-i\epsilon\int d^{4}x(\frac{1}{2}A^{2}-\overline{c}c)$}
\textmd{\emph{inside the $\kappa$-integral.}} \textmd{Thus, taking
care of the proper definition of the axial Green's functions, these
Green's functions are given in the present formalism by \cite{ijmpa00,jmp00}} }

\begin{eqnarray}
\left\langle \left\langle O[\phi]\right\rangle \right\rangle _{A} & = & \left\langle \left\langle O[\phi]\right\rangle \right\rangle _{L}\nonumber \\
 & + & i\int_{0}^{1}d\kappa\int D\phi exp\left\{ iS_{eff}^{M}[\phi,\kappa]+\epsilon\int d^{4}x(\frac{1}{2}A^{2}-\overline{c}c)\right\} \nonumber \\
 & \bullet & \sum_{i}(\widetilde{\delta_{1i}}[\phi]+\kappa\widetilde{\delta_{2i}}[\phi])(-i\Theta'\frac{\delta^{L}O}{\delta\phi_{i}})\label{basic}\end{eqnarray}
where $\Theta'$ has been determined in Ref. \cite{bj98,jm95} and
is given \begin{equation}
\Theta'[\phi]=i\int d^{4}y\,\overline{c}^{\gamma}(y)\left[\partial.A^{\gamma}(y)-\eta.A^{\gamma}(y)\right]\label{theta'}\end{equation}
and $S_{eff}^{M}[\phi,\kappa]$ is the effective action for the mixed
gauge function $F[A,\kappa]=(1-\kappa)\partial.A^{\gamma}(y)+\kappa\eta.A^{\gamma}(y)$
and $\widetilde{\delta_{1i}}[\phi]+\kappa\widetilde{\delta_{2i}}[\phi]$
stands for the BRS variation for $\phi$ for the mixed gauge \cite{jmp00}.
As discussed in Ref. \cite{jmp00}, a given Green's function to a
given finite order can be evaluated by means of a finite set of diagrams
with vertices from the Lorentz gauges and the BRS variations together
with the propagators and ghost-ghost-gauge vertex from the mixed gauges.
A $\kappa$-integral is also required to be performed.

For example, consider $O[\phi]=A_{\mu}^{\alpha}(x)A_{\nu}^{\beta}(y)$.
Then, $\langle\langle A_{\mu}^{\alpha}(x)A_{\nu}^{\beta}(y)\rangle\rangle_{A}=iG_{\mu\nu}^{A\alpha\beta}(x-y)$
for the connected part. Then, in obvious notations, (\ref{basic})
reads,\begin{eqnarray}
iG_{\mu\nu}^{A\alpha\beta}(x-y) & = & iG_{\mu\nu}^{L\alpha\beta}(x-y)\nonumber \\
 & + & \int_{0}^{1}d\kappa\int D\phi exp\left\{ iS_{eff}^{M}[\phi,\kappa]+\epsilon\int d^{4}x(\frac{1}{2}A^{2}-\overline{c}c)\right\} \nonumber \\
 & \bullet & [D_{\mu}c^{\alpha}(x)A_{\nu}^{\beta}(y)+A_{\mu}^{\alpha}(x)D_{\nu}c^{\beta}(y)]\nonumber \\
 & \bullet & \left.\int d^{4}z\overline{c}^{\gamma}(z)[\partial.A-\eta.A]^{\gamma}(z)\right|_{conn}\label{prop}\end{eqnarray}
The above relation gives the value of the exact axial propagator \emph{}in
the above formalism\emph{.} The result is exact to all orders. As
mentioned earlier, to any finite order in $g$, the right hand side
can be evaluated by a sum of a finite number of Feynman diagrams.
In this work, we shall discuss some of the further applications of
these results to the comparative renormalization in the axial and
the Lorentz gauges.

\subsection{Additional Feynman Rules}

For completeness, we shall give the mixed gauge propagators that enter
the Feynman diagrams that arise from $S_{eff}^{M}[\phi,\kappa]$ and
new vertices:

\begin{itemize}
\item Gauge boson propagator $iG_{\mu\nu}^{0M}(k,\kappa)$\begin{eqnarray}
 & G_{\mu\nu}^{0M}(k,\kappa)\nonumber \\
= & \frac{-1}{k^{2}+i\varepsilon}\left\{ g_{\mu\nu}+\frac{\mathcal{A}k_{\mu}k_{\nu}+\mathcal{B}\left(k_{\mu}\eta_{\nu}+k_{\nu}\eta_{\mu}\right)+\mathcal{C}\left(k_{\mu}\eta_{\nu}-k_{\nu}\eta_{\mu}\right)+\mathcal{E}\eta_{\mu}\eta_{\nu}}{\mathcal{D}}\right\} \label{gaugeprop}\end{eqnarray}
with\[
\mathcal{A}=(1-\kappa)^{2}-\frac{\eta^{2}\kappa^{2}}{k^{2}+i\varepsilon};\,\,\mathcal{B}=\frac{\kappa^{2}\eta.k}{k^{2}+i\varepsilon};\,\,\mathcal{C}=-i\kappa(1-\kappa);\,\,\mathcal{E}=\frac{i\kappa^{2}\varepsilon}{k^{2}+i\varepsilon}\]
\[
\mathcal{D}=\frac{-\kappa^{2}}{(k^{2}+i\varepsilon)}\left[(\eta.k)^{2}-\eta^{2}k^{2}+(k^{2}+\eta^{2})(k^{2}+i\varepsilon)\right]+2k^{2}\kappa-i\varepsilon\lambda-k^{2}\]

\item Ghost propagator $iG^{0M}(k,\kappa)$ (ghost carries momentum $k$)\begin{equation}
\tilde{{G}}^{0M}(k,\kappa)=\frac{1}{(\kappa-1)k^{2}-i\kappa k.\eta-i\varepsilon}\label{ghostprop}\end{equation}

\item $\bar{c}^{\alpha}c^{\beta}A_{\mu}^{\gamma}$vertex (antighost field
brings \emph{in} momentum $k$)\begin{equation}
-i\left[-ik_{\mu}(1-\kappa)+\kappa\eta_{\mu}\right]g_{0}f^{\alpha\beta\gamma}\label{ggA}\end{equation}

\item anti-ghost-gauge boson Vertex arising from $-i\Theta'$ (antighost
field brings \emph{in} momentum $k$):\begin{equation}
(-ik_{\mu}-\eta_{\mu})\delta^{\alpha\beta}\label{gAV}\end{equation}

\end{itemize}
For future reference, we note that for a 4-vector $e^{\mu}$ satisfying
$e.k=0=e.\eta$, the gauge boson propagator of (\ref{gaugeprop})
satisfies,\begin{equation}
e^{\mu}G_{\mu\nu}^{0M}(k,\kappa)=-\frac{e_{\nu}}{k^{2}+i\varepsilon}=a\,\, quantity\,\, independent\,\, of\,\,\kappa\label{eG}\end{equation}

\section{{\normalsize Relation Between Wavefunction Renormalization Constants
For the Axial and the Lorentz Gauges}}

The aim in this work is to develop a way to compare the calculations
and the renormalization procedure in the two gauges. With this in
view, in this section, we shall derive, as an application of (\ref{prop}),
the relation between the wavefunction renormalization constants for
the axial and the Lorentz gauges. The derivation is based upon the
relation (\ref{basic}) that gives an arbitrary Green's function in
the axial gauges {[}with proper treatment of singularities automatically
included{]} in terms of the Green's functions in the Lorentz gauges.
In the next section, we shall address to the calculation of the the
3-point function and in particular, the $\beta$-function.

We shall start with the relation (\ref{basic}) as applied to the
2-point function viz. that given by (\ref{prop}) in one loop approximation.
In doing so, we need to keep terms of $O(g^{2})$ in the equation
(\ref{prop}). 

We note that to extract the wavefunction renormalization constant
in the axial gauges from this relation, we need to obtain only those
terms in this relation that are proportional to $g_{\mu\nu}$. We
start from Eq.(\ref{prop}):\begin{eqnarray}
iG_{\mu\nu}^{A\alpha\beta}(x-y)\nonumber \\
=iG_{\mu\nu}^{L\alpha\beta}(x-y) & + & \int_{0}^{1}d\kappa\langle\langle[D_{\mu}c^{\alpha}(x)A_{\nu}^{\beta}(y)]\nonumber \\
 & \bullet & \int d^{4}z\overline{c}^{\gamma}(z)[\partial.A-\eta.A]^{\gamma}(z)\nonumber \\
 & + & a\,\, Bose-symmetric\,\, term\rangle\rangle_{conn}\label{prop'}\end{eqnarray}
in the one loop approximation. The second term on the right hand side
can be evaluated by drawing all one-loop diagrams with one insertion
each of the composite operators $D_{\mu}c^{\alpha}(x)$ and $\int d^{4}z\overline{c}^{\gamma}(z)[\partial.A-\eta.A]^{\gamma}(z)$;
we need to take only the connected parts of these terms. These diagrams
are functions of $\kappa$ , and a $\kappa$-integral is to be performed
in the end. These diagrams are as shown in Fig.1 {[}fig. (1A)-(1D){]}.
There are also 4 more diagrams obtained by Bose symmetrization$(x,\alpha,\mu)\leftrightarrow(y,\beta,\nu)$.

\includegraphics[%
  scale=0.5]{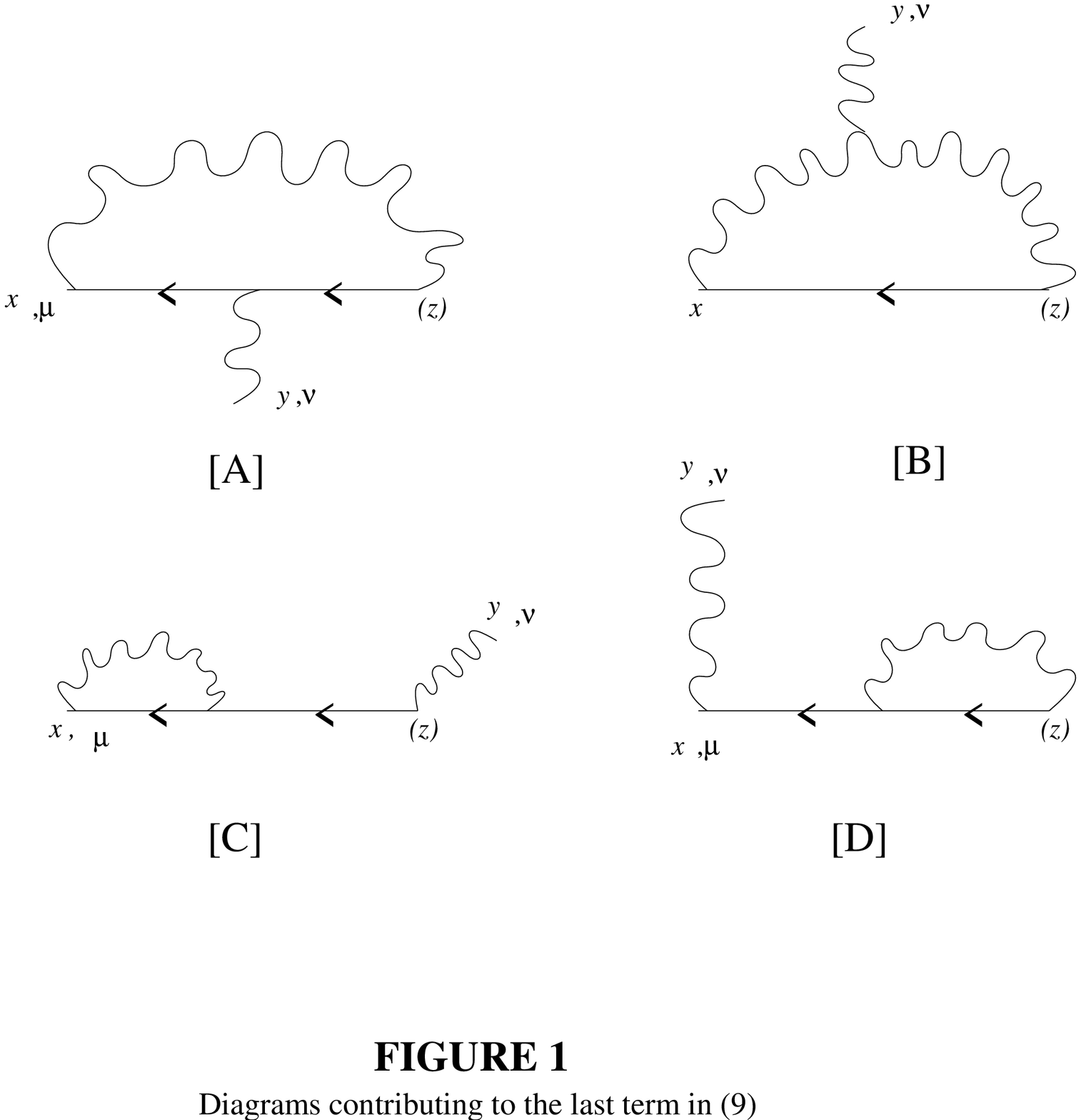}

We consider the relation (\ref{prop'}) in the momentum space. Let
$p_{\mu}$ be the 4-momentum of the gauge boson entering at $x$.
Then each term in (\ref{prop'}) has the form,\begin{equation}
g_{\mu\nu}A'+\eta_{\mu}p_{\nu}B'+\eta_{\nu}p_{\mu}C'+p_{\mu}p_{\nu}D'+\eta_{\mu}\eta_{\nu}E'\label{form}\end{equation}
 where $A',......E'$ are functions of $p^{2},\eta^{2},\eta.p$ .
To find the relation between the wave function renormalization constants
in the two sets of gauges, it is sufficient to pick up terms proportional
to $g_{\mu\nu}$ in each of the three terms in (\ref{prop'}). These
are easily projected from this relation by contracting it with a 4-vector
$e^{\mu}$ with the properties that $\mathcal{\textrm{$e.p$$=0=e.\eta$}}$
and $e^{2}\neq0$.

Now, we express \begin{equation}
Div\left\{ e^{\mu}e^{\nu}G_{\mu\nu}^{L\alpha\beta}(p,\eta)\right\} =-\delta^{\alpha\beta}\frac{Z_{L}}{p^{2}+i\varepsilon}e^{2}.\label{ZL}\end{equation}
We note that the digram 1(C) has terms that are necessarily proportional
to either $p_{\mu}$ or $\eta_{\mu}$ and hence do not contribute
on contraction with $e^{\mu}$; a similar statement holds for the
diagram obtained by Bose-symmetrization. The diagram (1D) vanishes
on account of the antisymmetry of the structure constants {[}we assume
a semi-simple gauge group{]}. We thus need to consider only the diagrams
1(A) and 1(B). We shall call the 1PI parts of these digrams {[}i.e.
the ones obtained by setting aside the gauge propagator $G_{\mu\nu}^{0M}$
{]} as $D_{\mu\sigma}^{A}$ and $D_{\mu\sigma}^{B}$. The tensorial
structure of each of these is of the form of (\ref{form}). The net
contribution from these diagrams reads 

\begin{equation}
i\int_{_{0}}^{1}d\kappa\left\{ \left[D_{\mu\sigma}^{A}+D_{\mu\sigma}^{B}\right]G_{\sigma\nu}^{0M}(p)+terms\,\, obtained\,\, by\,\, Bose-symmetrization\right\} \label{contr}\end{equation}
 We consider the $[D_{\mu\sigma}^{A}+D_{\mu\sigma}^{B}]$ and note
that it is a dimensionless tensor of rank two constructed out of $g_{\mu\nu},p_{\mu},\eta_{\mu}$.
Moreover, as $\lambda\rightarrow0$, the theory is invariant under
a uniform scaling $\eta_{\mu}\rightarrow\xi\eta_{\mu}$. Hence, we
parametrize: 

\begin{equation}
[D_{\mu\sigma}^{A}+D_{\mu\sigma}^{B}]=\frac{i}{2}\left\{ Bg_{\mu\sigma}+A\frac{p_{\mu}p_{\sigma}}{p^{2}}+C\frac{p_{\mu}\eta_{\sigma}}{\eta.p}+D\frac{p_{\sigma}\eta_{\mu}}{\eta.p}+E\frac{\eta_{\mu}\eta_{\sigma}}{\eta^{2}}\right\} \label{div}\end{equation}
where $A,B,C,D,E$ are dimensionless scalar functions of $\eta^{2},p^{2},\eta.p$
and $\kappa$. In fact, since there is only one scalar that is invariant
under scaling of $\eta$, viz. $\frac{\left(\eta.p\right)^{2}}{\eta^{2}p^{2}}$,
they are all functions (possibly) of this variable. When (\ref{div})
is substituted in (\ref{contr}) and is contracted with $e^{\mu}e^{\nu}$,
only the $g_{\sigma\nu}$ type terms in each of $D_{\mu\sigma}$ and
$G_{\sigma\nu}^{0M}$ contribute. Then, the net result for the diagrams
1(A) and 1(B) are given by,\begin{eqnarray}
\left[1(A)+1(B)\right] & = & i\int_{_{0}}^{1}d\kappa\left[D_{\mu\sigma}^{A}+D_{\mu\sigma}^{B}\right]G_{\sigma\nu}^{0M}(p)e^{\mu}e^{\nu}\nonumber \\
 & + & terms\,\, obtained\,\, by\,\, Bose-symmetrization\nonumber \\
 & = & \int_{_{0}}^{1}d\kappa B\left(\epsilon,\kappa,\frac{\left(\eta.p\right)^{2}}{\eta^{2}p^{2}}\right)\frac{_{1}}{p^{2}+i\varepsilon}e^{2}\nonumber \\
Div\left[1(A)+1(B)\right] & =Div & \int_{_{0}}^{1}d\kappa B\left(\epsilon,\kappa,\frac{\left(\eta.p\right)^{2}}{\eta^{2}p^{2}}\right)\times\frac{_{1}}{p^{2}+i\varepsilon}e^{2}\label{div1}\end{eqnarray}
 As confirmed by the calculations in \cite{jm-01}, the divergence
above is a local quantity in co-ordinate space (i.e. independent of
$\frac{\left(\eta.p\right)^{2}}{\eta^{2}p^{2}}$), so that \begin{equation}
Div\int_{_{0}}^{1}d\kappa B\left(\epsilon,\kappa,\frac{\left(\eta.p\right)^{2}}{\eta^{2}p^{2}}\right)=\hat{B}(\varepsilon,\eta).\label{div'}\end{equation}
Further, with this input, we can parametrize\begin{equation}
Div\left\{ e^{\mu}e^{\nu}G_{\mu\nu}^{A\alpha\beta}(p,\eta)\right\} =-\delta^{\alpha\beta}\frac{Z_{A}}{p^{2}+i\varepsilon}e^{2}.\label{ZA}\end{equation}
Substituting (\ref{ZA}),(\ref{ZL}) and (\ref{div1}) in (\ref{prop'}),
we obtain 

\begin{equation}
Z_{A}=Z_{L}-\hat{B}(\epsilon,\eta)\label{ZAZL}\end{equation}

\section{{\normalsize Gauge independence of the $\beta$-function}}

In this section, we shall compare the evaluation of the 3-point function
in the two gauges and shall show how the relation (\ref{basic}),
when applied to the three-point function, leads to the equality of
the $\beta$-functions in the axial gauges and the Lorentz gauges
in the one loop approximation. We iterate that the proof proceeds
by rigorous formal arguments which do not ignore the essential problem
of axial gauges: the pole treatment.

We consider the relation (\ref{basic}) for the case when 

\begin{equation}
O=A_{\mu}^{\alpha}(x)A_{\nu}^{\beta}(y)A_{\sigma}^{\gamma}(z)\label{O}\end{equation}
The equation (\ref{basic}) for this case reads,

\begin{eqnarray}
 & G_{\mu\nu\sigma}^{A\alpha\beta\gamma}(x,y,z)-G_{\mu\nu\sigma}^{L\alpha\beta\gamma}(x,y,z)\nonumber \\
= & i\int_{0}^{1}d\kappa\int D\phi exp\left\{ iS_{eff}^{M}[\phi,\kappa]+\epsilon\int d^{4}x(\frac{1}{2}AA-\overline{c}c)\right\} \nonumber \\
 & \bullet[D_{\mu}c^{\alpha}(x)A_{\nu}^{\beta}(y)A_{\sigma}^{\gamma}(z)+Bose-symmetric\,\, terms]\nonumber \\
 & \times\int d^{4}w\overline{c}^{\gamma}(w)[\partial.A^{\gamma}(w)-\eta.A^{\gamma}(w)]\nonumber \\
\equiv & i\int_{_{0}}^{1}d\kappa\langle\langle[D_{\mu}c^{\alpha}(x)A_{\nu}^{\beta}(y)A_{\sigma}^{\gamma}(z)+Bose\, symmetric\, terms]\Theta'\rangle\rangle_{\frac{conn}{mixed}}\label{vert}\end{eqnarray}
We shall consider (\ref{vert}) in the momentum space. Let $p,q,r$
be respectively the three momenta $p+q+r=0$. We shall choose the
momenta and the two polarization vectors e and e' with the properties
$e.p=0=e.q=e.\eta=e.r$; and $e'.r=e'.\eta=0$. We shall assume that
$\eta$ and the momenta under consideration $p,q$ are linearly independent
4-vectors. Let us now consider $e^{\mu}e^{\nu}e'^{\sigma}G_{\mu\nu\sigma}^{A\alpha\beta\gamma}(p,q,\eta)$.
Considering the fact that $e^{\mu}e^{\nu}G_{\mu\nu\sigma}^{A\alpha\beta\gamma}(p,q,\eta)$
is quadratic in $e$ and has one free index $\sigma$, this quantity
has the following tensors appearing in its expansion:

\begin{equation}
e^{2}p_{\sigma},e^{2}q_{\sigma},e^{2}\eta_{\sigma},e.pe_{\sigma},e.qe_{\sigma},and\,\, e.\eta e_{\sigma}\label{terms}\end{equation}
 Of these, the last three tensors vanish identically. Further, on
account of the fact that \[
e^{2}\eta_{\sigma}e'^{\sigma}=0=e^{2}(p_{\sigma}+q_{\sigma})e'^{\sigma};\]
 $e^{\mu}e^{\nu}e'^{\sigma}G_{\mu\nu\sigma}^{A\alpha\beta\gamma}(p,q,\eta)$
can be expressed as

\begin{equation}
e^{\mu}e^{\nu}e'^{\sigma}G_{\mu\nu\sigma}^{A\alpha\beta\gamma}(p,q,\eta)=e^{2}e'.(p-q)\hat{A}(p^{2},q^{2},p.q,\eta^{2},\eta.p,\eta.q,\epsilon)\frac{1}{p^{2}q^{2}r^{2}}f^{\alpha\beta\gamma}g_{0}\label{GA3}\end{equation}
$G_{\mu\nu\sigma}^{A\alpha\beta\gamma}(p,q,\eta)$ involves three
external bare propagators, which together with factors $e^{\mu}e^{\nu}e'^{\sigma}$
lead to an explicit dependence $\frac{1}{p^{2}q^{2}r^{2}}$. Thus%
\footnote{Below, by divergence we shall mean the divergence in the unrenormalized
Green's function in 1-loop approximation that arises from the \emph{loop
integrations.}%
},

\begin{eqnarray}
Div[e^{\mu}e^{\nu}e'^{\sigma}G_{\mu\nu\sigma}^{A\alpha\beta\gamma}(p,q,\eta)]\nonumber \\
=g_{0}f^{\alpha\beta\gamma}e^{2}e'.(p-q)Div\left[\widehat{A}(p^{2},q^{2},p.q,\eta^{2},\eta.p,\eta.q,\epsilon)\right]\frac{1}{p^{2}q^{2}r^{2}}\label{divA}\end{eqnarray}
A similar argument leads us further to

\begin{equation}
Div[e^{\mu}e^{\nu}e'^{\sigma}G_{\mu\nu\sigma}^{L\alpha\beta\gamma}(p,q)]=g_{0}f^{\alpha\beta\gamma}e^{2}e'.(p-q)\widehat{L}(\epsilon)\frac{1}{p^{2}q^{2}r^{2}}\label{divL}\end{equation}
(where the locality of the divergence in Lorentz gauges is already
known) and

\begin{eqnarray}
Div[e^{\mu}e^{\nu}e'^{\sigma}R_{\mu\nu\sigma}^{\alpha\beta\gamma}(p,q,\eta)]\nonumber \\
=g_{0}f^{\alpha\beta\gamma}e^{2}e'.(p-q)Div\left[\widehat{R}(p^{2},q^{2},p.q,\eta^{2},\eta.p,\eta.q,\epsilon)\right]\frac{1}{p^{2}q^{2}r^{2}}\label{divR}\end{eqnarray}
where the $R_{\mu\nu\sigma}^{\alpha\beta\gamma}(p,q,\eta)$ stands
for the contribution from the last term in (\ref{vert}). 

The renormalization constant for the coupling $g_{0}$ in 1-loop approximation
can be extracted by locating the constant divergence in the residue
of $e^{\mu}e^{\nu}e'^{\sigma}G_{\mu\nu\sigma}^{A\alpha\beta\gamma}(p,q,\eta)$
near its poles at $p^{2}=0,q^{2}=0,\,\, and\,\, r^{2}=0$ (Please
see relations (\ref{AZ1Z})).

To find the contributions for $R_{\mu\nu\sigma}^{\alpha\beta\gamma}(p,q,\eta)$,
we need to find the Feynman diagrams for this term.These are as shown
in Fig.2, viz. \{\emph{fig(2a)--fig(2o)}\}. With each of these, there
are also the digrams obtained by appropriate permutations. Despite
a large number of these diagrams, only a few contribute to $Div[\widehat{R}]$
of (\ref{divR}). 

\begin{itemize}
\item \includegraphics[%
  scale=0.5]{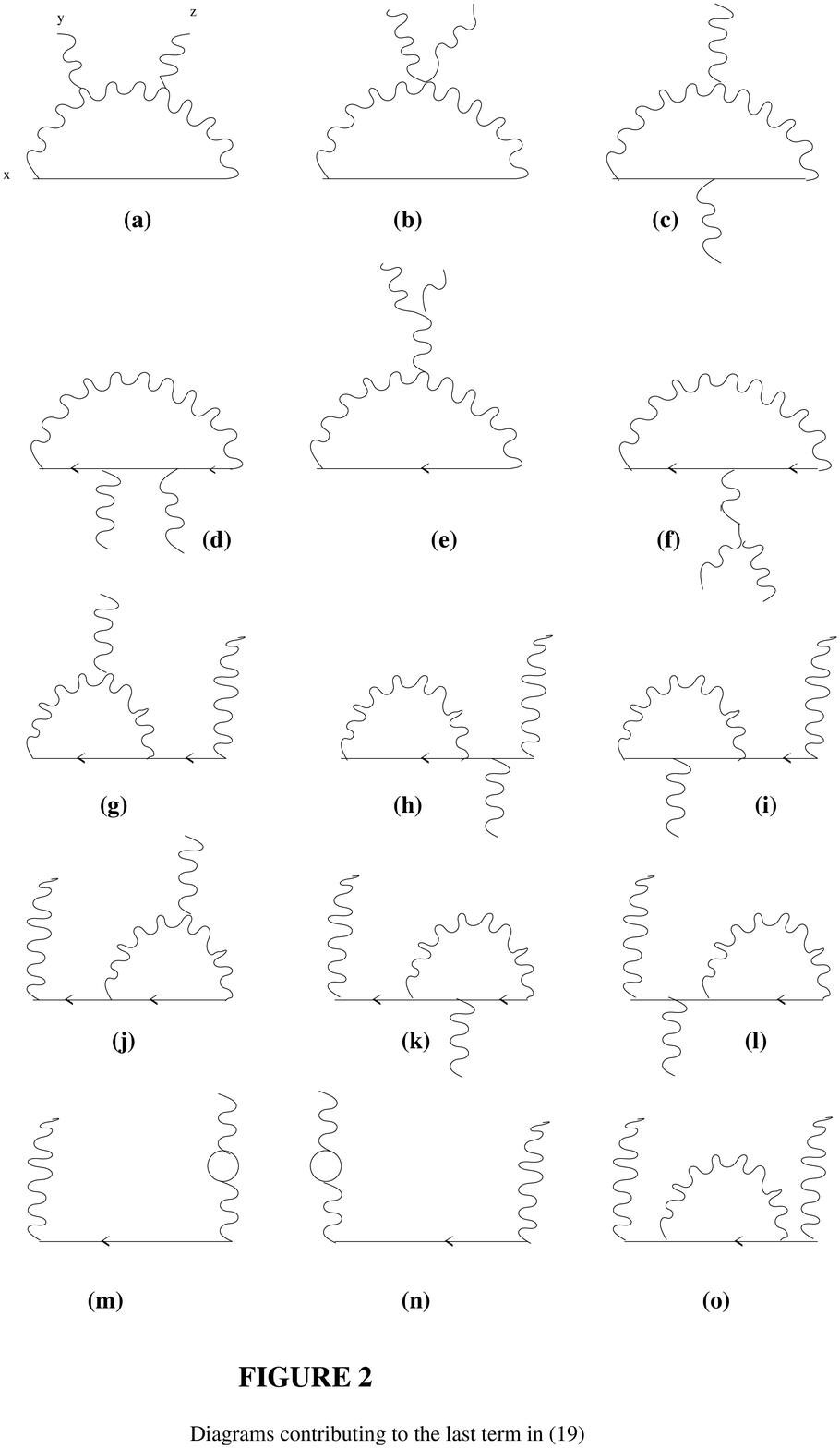}
\end{itemize}
We shall now proceed to show that as far as the coupling constant
renormalization constant is concerned, only the diagrams \emph{2(e)}
and \emph{2(f)} need to be taken into account and that the divergence
in these is related to that in the diagrams 1(A) and 1(B) taken together
(See (\ref{div1})). There are more than one ways to deal with this.
Before we do this, however, we should recall that all the lines in
a graph are generally $\kappa$-dependent and an integral over $\kappa$
is to be done at the end.

\begin{itemize}
\item Diagrams \emph{(g),(h),(i),(n)} and \emph{(o)} have a ghost line followed
by a gauge-boson line with a vertex arising from (\ref{theta'}).
This gauge boson line is contracted with $\varepsilon_{\mu}$ the
polarization vector of that gauge-boson (which is either $e$ or $e'$
depending on the context). The vertex is of the form $\alpha\eta_{\sigma}+\beta s_{\sigma}$
where $\alpha$ and $\beta$ are some constants and $s$ is the momentum
of the gauge-boson (See Eq.(\ref{gAV})). Then, these diagrams vanish
because of $\left(\alpha\eta_{\sigma}+\beta s_{\sigma}\right)\varepsilon^{\sigma}=0$.
The same logic also holds for \emph{(m)}.
\item Alternately, the diagram \emph{(h)} vanishes\emph{,} being proportional
to $(\gamma\eta_{\mu}+\delta p_{\mu})e^{\mu}\equiv0$.
\item Diagram \emph{(l)} vanishes by the anti-symmetry of the structure
constants.
\item Diagrams \emph{(j)} and \emph{(k)} have \emph{no poles} in $p^{2}$.
Similar diagrams obtained by permutations have no poles in the respective
$(momentum)^{2}$. The absence of pole in $p^{2}$ is also noted for
diagrams \emph{(m),(n)} and \emph{(o)} (which were already shown to
vanish by alternate means).
\item Diagrams \emph{(a),(b),(c)} and \emph{(d)} also have no poles in $p^{2}$.
This is seen by first noting that these diagrams exist at $p=0$ followed
by a simple kinematical analysis.
\end{itemize}
This leaves us with diagrams of \emph{2(e)} and \emph{2(f)} and their
permutations. These two diagrams add up to 

\[
i\int_{0}^{1}d\kappa[D_{\mu\rho}^{A}(p,\kappa)+D_{\mu\rho}^{B}(p,\kappa)]G_{\rho\nu\sigma}^{0M\alpha\beta\gamma}(p,q,r;\kappa)\]
Here $G_{\rho\nu\sigma}^{0M\alpha\beta\gamma}(p,q,r;\kappa)$ is the
bare three point Green's function in the mixed action, which consists
of the bare three-point vertex (which is independent of $\kappa$)
and three $\kappa$-dependent propagators. We contract this with $e^{\mu}e^{\nu}e'^{\sigma}$
and note relations such as (See (\ref{eG})) \[
e^{\mu}G_{\mu\nu}^{0M}(p,\eta,\kappa)=\frac{-e_{\nu}}{p^{2}+i\varepsilon}=a\,\,\kappa-independent\,\, expression\]
(and two analogous relations), the $\kappa$-dependence arising from
$G_{\rho\nu\sigma}^{0M\alpha\beta\gamma}(p,q,r;\kappa)$ drops and
we obtain, on simplification,\begin{eqnarray*}
 &  & ie^{\mu}e^{\nu}e'^{\sigma}\int_{0}^{1}d\kappa[D_{\mu\rho}^{A}(p,\kappa)+D_{\mu\rho}^{B}(p,\kappa)]G_{\rho\nu\sigma}^{0M\alpha\beta\gamma}(p,q,r;\kappa)\\
 & = & \int_{0}^{1}d\kappa[D_{\mu\rho}^{A}(p,\kappa)+D_{\mu\rho}^{B}(p,\kappa)]\times e^{2}\frac{e'.(p-q)}{p^{2}q^{2}r^{2}}f^{\alpha\beta\gamma}g_{0}\\
 & = & \frac{1}{2}\int_{_{0}}^{1}d\kappa B\left(\epsilon,\kappa,\frac{\left(\eta.p\right)^{2}}{\eta^{2}p^{2}}\right)\times e^{2}\frac{e'.(p-q)}{p^{2}q^{2}r^{2}}f^{\alpha\beta\gamma}g_{0}\end{eqnarray*}
We note that the above expression involves the same 1PI part of diagrams
as contained in the expression (\ref{contr}) to the 2-point function.
We recall that (\ref{div'}),

\[
Div\left[\frac{1}{2}\int_{_{0}}^{1}d\kappa B\left(\epsilon,\kappa,\frac{\left(\eta.p\right)^{2}}{\eta^{2}p^{2}}\right)\right]=\frac{1}{2}\hat{B}(\epsilon,\eta)\]
(We have already argued why this is a constant independent of $p$).
Thus,\begin{eqnarray*}
ie^{\mu}e^{\nu}e'^{\sigma}Div\left[\int_{0}^{1}d\kappa[D_{\mu\rho}^{A}(p,\kappa)+D_{\mu\rho}^{B}(p,\kappa)]G_{\rho\nu\sigma}^{0M\alpha\beta\gamma}(p,q,r;\kappa)\right]\\
=\frac{1}{2}\hat{B}e^{2}\frac{e'.(p-q)}{p^{2}q^{2}r^{2}}f^{\alpha\beta\gamma}g_{0}\end{eqnarray*}
 A similar result holds for diagrams obtained by permutations of $x,y,z$
labels as in (\ref{vert}) and thus leads to a net divergent contribution
$=\frac{3}{2}\hat{B}e^{2}\frac{e'.(p-q)}{p^{2}q^{2}r^{2}}f^{\alpha\beta\gamma}g_{0}$.
Then (\ref{vert}) leads us to 

\begin{eqnarray}
 & Div\left[\widehat{A}(p^{2},q^{2},p.q,\eta^{2},\eta.p,\eta.q,\epsilon)\right]_{_{p^{2}=q^{2}=r^{2}=0}}\nonumber \\
\equiv & \hat{A}(\epsilon,\eta)=\widehat{L}(\epsilon)+\frac{3}{2}\widehat{B}(\epsilon,\eta)\label{ALB}\end{eqnarray}
and thus is a constant. Now, we recall that the 1-loop diagrams for
the unrenormalized 3-point function consist of those with the vertex
modification or with a self-energy insertion. This implies,\begin{eqnarray}
\hat{A}(\epsilon,\eta) & = & (Z_{1A}-1)-3(Z_{A}-1)\nonumber \\
\widehat{L}(\epsilon) & = & (Z_{1L}-1)-3(Z_{L}-1)\label{AZ1Z}\end{eqnarray}
We combine (\ref{AZ1Z}) and (\ref{ALB}) with 

\[
Z_{A}=Z_{L}-\widehat{B}(\epsilon,\eta)\]
to find that up to 1-loop approximation,\[
(Z_{1A}-Z_{1L})-\frac{3}{2}(Z_{A}-Z_{L})=0\]
 This leads to $\textrm{$Z_{g,A}$$=Z_{g,L}$}$, where we have defined
$\textrm{$Z_{g,A}$$=\frac{Z_{1A}}{Z_{A}^{3/2}}$}$and \textrm{\textrm{$Z_{g,L}$$=\frac{Z_{1L}}{Z_{L}^{3/2}}$}}
, thus implying the identity of the beta functions in the two gauges
in one loop order.

\textbf{\emph{ACKNOWLEDGMENTS}}

I would like to acknowledge financial support from Department of Science
and Technology,  Government of India in the form of a grant for the
project No. DST-PHY-19990170.

\end{document}